\documentclass[twocolumn]{aastex631}
\usepackage{amsmath}
\usepackage{amssymb}
\usepackage{apjfonts}

\newlength{\apjcolwidth}
\begin{document}

\title{Carbon-Oxygen Phase Separation in MESA White Dwarf Models}

\correspondingauthor{Evan B. Bauer}
\email{evan.bauer@cfa.harvard.edu}

\author[0000-0002-4791-6724]{Evan~B.~Bauer}
\affiliation{Center for Astrophysics $\vert$ Harvard \& Smithsonian, 60 Garden St, Cambridge, MA 02138, USA}

\begin{abstract}
  We enhance the treatment of crystallization for models of white
  dwarfs (WDs) in the stellar evolution software MESA
  by implementing carbon-oxygen (C/O) phase
  separation. The phase separation process during crystallization
  leads to transport of oxygen toward the center of WDs,
  resulting in a more compact structure that liberates gravitational
  energy as additional heating that modestly slows WD cooling
  timescales. We quantify this cooling delay in MESA C/O WD models
  over the mass range $0.5-1.0\,M_\odot$, finding delays
  of $0.5-0.8\,\rm Gyr$ for typical C/O interior profiles. MESA WD cooling
  timescales including this effect are generally comparable to other
  WD evolution models that make similar assumptions about input
  physics. When considering phase separation alongside $^{22}$Ne
  sedimentation, however, we find that
  both MESA and BaSTI WD cooling models predict a more modest
  sedimentation delay than the latest LPCODE models,
  and this may therefore require a re-evaluation of
  previously proposed solutions to some WD cooling anomalies
  that were based on LPCODE models of $^{22}$Ne sedimentation.
  Our implementation of C/O phase separation in the open-source
  stellar evolution software MESA provides an important tool for
  building realistic grids of WD cooling models, as well as a framework
  for expanding on our implementation to explore additional physical
  processes related to phase transitions and associated fluid motions
  in WD interiors.
\end{abstract}


\section{Introduction}

White dwarf stars (WDs) cool over long timescales governed by the
thermodynamics of their interior heat reservoirs, mediated by the
physics of heat transport through their outer envelopes \citep{Mestel1952}.
With detailed WD models, these cooling timescales can be calculated
accurately and used to constrain the ages of individual WDs and
stellar populations that they are associated with
\citep{Winget1987,Fontaine2001}. Discrepancies between WD cooling ages
and independent age measurements can also be used to learn about the
physical process operating in the dense plasma mixtures of WD
interiors (e.g., \citealt{Garcia-Berro2010,Cheng2019,Bauer2020,Blouin2021apjl}).
Several WD evolution codes attempt to
implement the most up-to-date input physics and methods to provide
grids of models with accurate WD cooling timescales
(e.g., \citealt{Camisassa2016,Bedard2020,Bedard2022,Salaris2022,Jermyn2023}).

One important physical process that can modify WD cooling timescales is the
first-order phase transition of the carbon-oxygen (C/O) mixtures in WD
cores from liquid to solid when they cool to the point of crystallization
\citep{vanHorn1968,Winget2009}. The latent heat associated with this
phase transition can temporarily slow WD cooling, and phase separation
into O-enriched solid material and C-enriched liquid mantle material
can induce mixing that also impacts WD interior thermodynamics and
further delays WD cooling 
\citep{Stevenson1977,Mochkovitch1983,Segretain1993,Horowitz2010,Blouin2020}.

The physics of crystallization and C/O phase separation should be
included in detailed WD cooling models, and recent observations have
also motivated further investigation of other phenomenology that may
be tied to the physics of dense multi-component fluids.
The fluid motions in the liquid mantles of crystallizing WDs have
recently been proposed as a candidate mechanism for generating dynamos in
some magnetic WDs \citep{Isern2017,Ginzburg2022}. While a body of
observational evidence seems to support this proposal
\citep{Belloni2021,Schreiber2021,Schreiber2021mnras,Schreiber2022},
recent numerical simulations have called into question
whether the fluid motions are adequate to produce the necessary dynamo
effect \citep{Fuentes2023}.
The observed cooling anomaly for massive crystallizing WDs on the
``$Q$-branch'' \citep{Cheng2019,Bauer2020,Camisassa2021} has also
spurred recent investigation of previously unexplored phenomena
associated with impurities effecting the phase diagrams of dense
plasma mixtures
\citep{Caplan2020,Blouin2021apjl,Blouin2021apj,Horowitz2021,Caplan2023}.

This context motivates providing an implementation of C/O phase
separation in the open-source stellar evolution software instrument
MESA \citep{Paxton2011,Paxton2013,Paxton2015,Paxton2018,Paxton2019,Jermyn2023}
to aid in exploring the variety of phenomena associated with WD
crystallization and mantle mixing as new observations continue to
motivate more detailed investigations with stellar evolution models.
MESA already includes state-of-the-art capabilities for many aspects
of WD evolution, including the Skye EOS \citep{Jermyn2021} that
includes crystallization and latent heat, and $^{22}$Ne
sedimentation that can independently delay WD cooling
\citep{Paxton2018,Bauer2020} with the diffusion coefficients of
\cite{Caplan2022} that are accurate for dense, strongly-coupled liquid
plasmas. In this work, we further enhance the available physics
capabilities for MESA WD models by providing an implementation of C/O
phase separation based on the recent phase diagram of
\cite{Blouin2021}.

We begin by briefly describing the setup for our MESA models and
key pieces of input physics in Section~\ref{s.input}.
Section~\ref{s.implementation} then gives a detailed description of
our method for implementing C/O phase separation in MESA, and
shows how this implementation results in changes to interior
composition profiles and heating that delays WD cooling.
Section~\ref{s.cooling} shows the impact that C/O phase separation has
on WD cooling timescales in MESA models, and compares to other
commonly used WD cooling models.
Section~\ref{s.ne22} then explores the interplay between
cooling delays from both C/O phase separation and $^{22}$Ne
sedimentation in models descended from solar-metallicity progenitors
where both effects are significant. We find important differences
between MESA models and other commonly used WD cooling models when
including $^{22}$Ne sedimentation, and we suggest that our models
require new input physics to resolve important WD cooling
discrepancies such as the WD $Q$-branch and the WD luminosity function
of NGC~6791. This motivates further work to explore stellar evolution
models that include the physics of multi-component phase diagrams that
account for impurities altering the C/O phase separation process,
which our work here lays the groundwork for pursuing with MESA in the
future.

\section{Code and Input Physics}
\label{s.input}

Our MESA models presented in this work rely on the development version
of MESA, commit {\tt 59ed280}, which is publicly available on the
{\tt main} branch of the MESA GitHub repository.%
\footnote{\url{https://github.com/MESAHub/mesa}}
In most respects, this version functions very similarly to the most
recent public release of MESA as of the writing of this work,
r22.11.1. The C/O phase separation capabilities that are presented in
this work are available on the {\tt main} branch of the development
version of MESA and will therefore be available in the next public
release version of MESA after r22.11.1.
A repository of work directories containing MESA input files
needed to reproduce all of the models presented in this work is
available on Zenodo: \dataset[doi:10.5281/zenodo.7742475]{\doi{10.5281/zenodo.7742475}}.

Radiative opacities are primarily from OPAL \citep{Iglesias1993,
Iglesias1996}, with low-temperature data from \citet{Ferguson2005}
and the high-temperature, Compton-scattering dominated regime by
\citet{Poutanen2017}.  Electron conduction opacities are from
\citet{Cassisi2007} and \citet{Blouin2020apj}.
Nuclear reaction rates are from JINA REACLIB \citep{Cyburt2010}, NACRE \citep{Angulo1999} and
additional tabulated weak reaction rates \citet{Fuller1985, Oda1994,
Langanke2000}.  Screening is included via the prescription of \citet{Chugunov2007}.
Thermal neutrino loss rates are from \citet{Itoh1996}.

The MESA equation of state (EOS) is a blend of the OPAL \citep{Rogers2002}, SCVH
\citep{Saumon1995}, FreeEOS \citep{Irwin2004}, HELM \citep{Timmes2000},
PC \citep{Potekhin2010}, and Skye \citep{Jermyn2021} EOSes. In
particular, the Skye EOS covers the region of high density where
crystallization occurs in WD interiors. Skye includes the latent heat
generated by crystallizing material \citep{Jermyn2021}, and since the
thermodynamics account for the full composition vector, Skye can also be
used to capture the heating associated with composition changes in a
WD interior during the phase separation process, as we will describe
in detail in later sections.

\section{Phase Separation Implementation}
\label{s.implementation}

For a crystallizing C/O mixture in a WD interior, the boundary between
liquid and solid has a discontinuous composition profile, with
the solid interior enriched in O relative to the surrounding C/O
liquid. A fluid element undergoing this transition minimizes the free
energy by separating into O-enriched solid and C-enriched liquid
components that are in equilibrium at the phase boundary
\citep{Stevenson1977,Segretain1993,Horowitz2010,Medin2010,Blouin2020}.
This extrusion of C into the liquid immediately surrounding the
crystallized core then excites fluid instability that
leads to mixing in the liquid layers, which releases heat associated
with the changing composition profile and slows the cooling of the WD,
with an overall delay on the order of up to $\approx$1~Gyr
\citep{Mochkovitch1983,Isern1997,Montgomery1999,Althaus2012}.
In this section, we provide a detailed description of the steps
we have implemented in MESA to account for these effects and their
influence on WD cooling timescales.

\subsection{Phase Diagram at the Boundary}

The phase transition occurs when the temperature falls to a point
where the free energy of the solid phase is lower than the free energy
of the liquid phase. Skye calculates all thermodynamics
self-consistently from derivatives of the Helmholtz free energy $F$,
and therefore has direct access to the free energy information needed
to evaluate where the phase transition occurs in a C/O plasma
mixture \citep{Jermyn2021}. Skye evaluates the free energy for both
liquid and solid phases, and selects the phase that minimizes the free
energy. The liquid-solid phase boundary is a first-order phase
transition, with continuous free energy at the boundary but
discontinuous derivatives resulting in discontinuities in
thermodynamic quantities such as entropy
$s = -(\partial F/ \partial T)_\rho$.

The latent heat of
crystallization is encoded in the entropy discontinuity at the phase
boundary, but a sharp discontinuity is difficult to capture with the
numerical methods of stellar evolution codes such as MESA.
Skye therefore introduces a narrow smoothing of the entropy
discontinuity at the phase transition to capture the latent heat
in a continuous manner convenient for inclusion in stellar evolution
codes. Skye accomplishes this by defining a continuous {\tt phase}
parameter $\phi$ constructed as follows. Denote the free energy of the
ion mixture as $f_{\rm i} = \min(f_{\rm i,liquid},f_{\rm i,solid})$,
where $f$ is given in dimensionless units of free energy per
$k_{\rm B}T$ per ion. With the difference between the free energy of
the liquid phase and solid phase $\Delta f \equiv f_{{\rm i, liquid}}
- f_{\rm i, solid}$, the {\tt phase} parameter is then constructed as
\begin{equation}
  \phi = \frac{e^{\Delta f/w}}{e^{\Delta f/w}+1}~,
  \label{eq:phi}
\end{equation}
where $w = 10^{-2}$ is chosen as the blurring parameter to introduce a
narrow but continuous {\tt phase} variable that transitions from 0 to
1 when $f_{\rm i}$ is within a few percent of its value at the
location of the phase transition.
The phase transition from liquid to solid is located at $\phi = 0.5$,
with $\phi < 0.5$ being liquid material and $\phi > 0.5$ being solid
material. The latent heating term can then be constructed from a
continuous smoothed entropy evaluated by taking the derivative of
\begin{equation}
  f_{\rm i,smooth} \equiv (1-\phi) f_{\rm i,liquid} + \phi f_{\rm i,solid}~.
  \label{eq:fsmooth}
\end{equation}
For more details see section~3 of \cite{Jermyn2021}.%
\footnote{Note that equation~(41) in \cite{Jermyn2021} has a typo in
  its  definition of $f_{\rm i,smooth}$, which is why it does not
  agree exactly with Eqn~\eqref{eq:fsmooth} here.}

We therefore use the thermodynamics and phase information reported by
the Skye EOS in our MESA models to determine when the
fluid in a cooling WD has reached the point of crystallization, along
with the resulting latent heat released by the ensuing phase
transition. In terms of the plasma coupling parameter
$\Gamma \equiv e^2 \langle Z^{5/3}\rangle/a_e k_{\rm B} T$
(where $Z$ is ion charge and $a_e$ is average electron spacing),
\cite{Jermyn2021} show that the crystallization in a C/O
mixture according to Skye occurs around $\Gamma \approx 200-230$
depending on C/O ratio,
generally in good agreement with other state-of-the-art C/O phase
curves such as those of \cite{Horowitz2010}, \cite{Blouin2020}, and
\cite{Blouin2021}.
For our models throughout this paper, we specifically use Skye with
the \cite{Ogata1993} option for the solid mixing term of the free
energy, along with the options to extrapolate the thermodynamic fits
near the phase transition. This gives a crystallization temperature
within a few percent of the liquidus of the \cite{Blouin2020} phase
diagram for C/O mixtures (figure~8 of \citealt{Jermyn2021}, ``fits
extended'' lower panel).

In principle, Skye could also be used to solve the double-tangent
optimization problem for separating the C/O composition into
the solid and liquid components that minimize the free energy
\citep{Medin2010}. However, for the composition changes due to phase
separation at the boundary, we instead elect to use the
phase diagram of \cite{Blouin2021}. Specifically,
we use the fitting form given by equation~(34) and Table~II of that
work.

\begin{figure}
  \centering
  \includegraphics[width=\apjcolwidth]{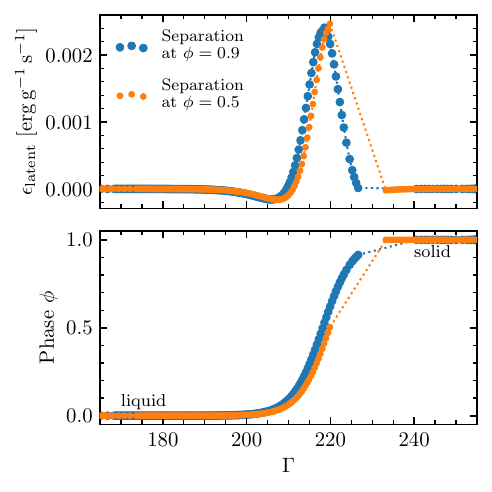}
  \caption{Latent heat and phase profiles as a function of plasma
    coupling $\Gamma$ for WD models undergoing both crystallization
    and phase separation using the Skye EOS. Points represent zones in
    the discretized stellar model, connected by thin dotted lines showing where
    the phase separation composition discontinuity lies.}
  \label{fig:Phase}
\end{figure}

The temperature profile in the core of a cooling WD is nearly
isothermal, so crystallization starts at the center where $\rho$ and
$\Gamma$ are highest. We begin the process of phase separation when
the model cools to a point where the phase in the central zone of the
model passes $\phi > \phi_{\rm sep}$. While the first-order phase
transition formally occurs precisely at $\phi = 0.5$ by construction
(Eqn~\ref{eq:phi}, $\Delta f=0$), we offset the location of the phase separation
induced composition discontinuity slightly further into the solid
phase by choosing $\phi_{\rm sep} = 0.9$ to avoid having phase
separation interfere with the release of the smoothed latent heat
term. This is because the method
for including the latent heat in Skye relies on density and
temperature evolution, and does not account for composition changes
\citep{Jermyn2021}.
For an individual fluid element to release all the
latent heat associated with evolving through the phase transition in
Skye, it needs to evolve continuously through $\phi$ near 0.5. Phase
separation causes a composition discontinuity that also causes $\phi$
to evolve discontinuously, as seen in  Figure~\ref{fig:Phase}.
This figure shows the latent heat in MESA WD models undergoing both
crystallization and phase separation according to our procedure using
the Skye EOS. Fluid elements evolve continuously from left to right in
this figure as they cool at nearly constant density, except where the
composition discontinuity from phase separation causes a discontinuous
jump to higher $\phi$ and $\Gamma$ shown by the dotted lines in the
figure.

The latent heat is strongly peaked at the location of
the phase transition from liquid to solid at $\phi = 0.5$,
which occurs at $\Gamma \approx 218$ in the C/O WD model used for
Figure~\ref{fig:Phase} according to Skye.
In the model where we impose phase separation precisely at
$\phi=0.5$, the composition discontinuity at that location causes fluid
elements to skip discontinuously past some of the latent heating near
the peak, causing some heat to be missed in the model. For the model
where we slightly delay phase separation to $\phi=0.9$,
the composition discontinuity is slightly offset into the solid phase at
$\Gamma \approx 226$ and therefore does not impact the latent heat
distribution in this model.
Offsetting the composition changes due to phase separation toward the
solid side of the phase transition allows each fluid element first to
release the smoothed latent heat of crystallization and then undergo
phase separation. 

\begin{figure}
  \centering
  \includegraphics[width=\apjcolwidth]{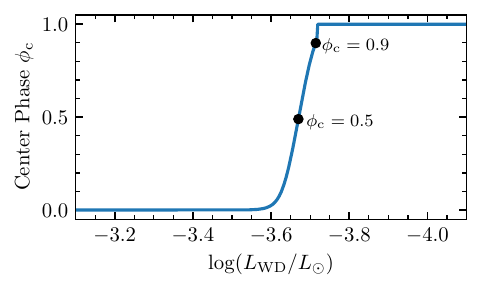}
  \caption{Evolution of the center phase $\phi_{\rm c}$ as a function
    of WD luminosity for a cooling and crystallizing WD model.}
  \label{fig:LumPhase}
\end{figure}

A fluid element evolves through $\phi = 0.9$
very soon after $\phi = 0.5$ because of the narrow blurring parameter
$w$, so any heat release associated with phase separation will occur
with only minimal delay and will not effect any resulting cooling
delay introduced into the overall WD evolution.
Figure~\ref{fig:LumPhase} shows that in the center of a crystallizing
WD model, $\phi$ evolves from 0.5 to 0.9 while the luminosity of the
WD changes by only about $\Delta \log L_{\rm WD} \approx 0.04$, or
about 10\%. This ensures that any heating associated with phase
separation and associated mixing will occur only slightly later due to
moving phase separation to $\phi = 0.9$. The slightly lower luminosity
at which the phase separation energy is released means that a
cooling delay could be overestimated, but by no more than about
10\%. Greater precision would simply require higher resolution models
and an even smaller blurring parameter $w < 10^{-2}$ to make the
smoothed phase transition and latent heat peak more strongly about the
location of $\phi = 0.5$, so that $\phi_{\rm sep}$ could be moved
closer to the location of the phase transition at $\phi=0.5$.

To summarize, as a mass coordinate in a WD interior cools and evolves
toward crystallization, we determine when it has reached the point of
crystallization from the thermodynamic and phase information reported
by the Skye EOS \citep{Jermyn2021}, and we adjust the composition at
the phase boundary using the C/O phase separation diagram of
\cite{Blouin2021}.

\subsection{Separation, Fluid Instability, and Mixing}

Here we describe the iterative procedure for propagating the
crystallization front and associated phase separation discontinuity
outward from the center of the model as the WD cools, along with the
associated mixing in the liquid mantle immediately surrounding the
solid core.

We label the discrete grid cells of a MESA model by zone number
$k=1,...,N_z$, with $k=1$ for the surface zone and $k = N_z$ for the
central zone (so that zone $k-1$ is the zone immediately outward from
zone $k$). We track the mass coordinate $m_{\rm cr}$ of the boundary
between the crystallized solid material and surrounding liquid. We assume
that this starts at $m_{\rm cr} = 0$ where density is highest in a WD,
and $m_{\rm cr}$ then evolves monotonically toward larger values as the WD cools. We
update and save this value between evolution steps to track which material is
newly crystallized (and therefore undergoing phase separation) in a
given time step. With $m_{\rm cr}$ therefore encoding the last known
location of the solid core boundary, we iterate outward starting from
the core boundary by starting at the largest $k$ such that the mass
coordinate $m_k$ of that zone satisfies $m_k > m_{\rm cr}$.
If $\phi_k>\phi_{\rm sep}$, then zone $k$ must undergo C/O phase
separation during the time step, so we adjust its composition by
enriching the oxygen mass fraction in zone $k$ by
\begin{equation}
  X_{{\rm O},k} \rightarrow X_{{\rm O},k} + \Delta X_{{\rm O},k}~,
\end{equation}
where $\Delta X_{{\rm O},k}$ is calculated as a function of $X_{{\rm O},k}$
using the fit of \cite{Blouin2021}. Conservation of mass requires that
we also adjust the carbon mass fraction in zone $k$ by
\begin{equation}
  X_{{\rm C},k} \rightarrow X_{{\rm C},k} - \Delta X_{{\rm O},k}~.
\end{equation}
This enrichment of O in zone $k$ extrudes C into zone $k-1$ to
compensate and conserve elements. Since adjacent cells may span
different amounts of mass depending on the mesh scheme of the model,
we must account for the mass $\delta m_k$ contained in zone $k$
relative to the mass $\delta m_{k-1}$ contained in zone $k-1$ to
properly conserve the amount of C/O mixed between the two cells. The mass
fractions of C and O in zone $k-1$ must therefore be adjusted by
\begin{equation}
  X_{{\rm C},k-1} \rightarrow
  X_{{\rm C},k-1} + \Delta X_{{\rm O},k} \delta m_k / \delta m_{k-1}~,
  \label{eq:XCout}
\end{equation}
and
\begin{equation}
  X_{{\rm O},k-1} \rightarrow
  X_{{\rm O},k-1} - \Delta X_{{\rm O},k} \delta m_k / \delta m_{k-1}~.
  \label{eq:XOout}
\end{equation}
After making these composition adjustments, we mark zone $k$ as having
crystallized by advancing $m_{\rm cr}$ to the mass coordinate of the
outer edge of zone $k$ ($m_{\rm cr} \rightarrow m_{\rm cr} + \delta m_k$).

Note that the C/O phase curves used for phase separation are
calculated assuming pure C/O mixtures, so for a WD model with other
trace metals present (e.g. Ne) we simplify by assuming that the other
traces are inert and unaffected at the phase boundary.
We rescale the C and O mass fractions up proportionally
such that $X_{\rm C/O} = X_{\rm C} + X_{\rm O} = 1$,
perform the phase separation composition adjustments at the boundary,
and then scale the compositions back down to reach the original
value of $X_{\rm C/O}$, while leaving the mass fractions of other
trace elements unchanged.

After advancing the core boundary through zone $k$, the fluid in the
surrounding zone $k-1$ is generally depleted of heavier O and enriched
in C relative to its surroundings according to
Eqns~\eqref{eq:XCout}--\eqref{eq:XOout}, leaving the model with an
inverted molecular weight gradient just outside the core boundary.
This can excite dynamical instability that mixes the liquid outward
\citep{Mochkovitch1983,Ginzburg2022}. This dynamical mixing takes
place on a timescale much shorter than the stellar evolution time
steps that we take for WD cooling, so we approximate this mixing as an
efficient process that fully mixes material outward until the
composition profile re-establishes stability,
following a procedure similar to ``convective pre-mixing'' described
in \cite{Paxton2019}.

We quantify instability leading to dynamical mixing in terms of the Ledoux
criterion for convection
\begin{equation}
  \nabla_T > \nabla_{\rm ad} + B~,
  \label{eq:Ledoux}
\end{equation}
where $\nabla_{\rm ad}$ is the local adiabatic temperature gradient
reported by the EOS, $B \propto \nabla_\mu$ is the Ledoux term accounting for the
composition gradient as constructed in \cite{Paxton2013} ($\nabla_\mu$ is the
molecular weight gradient), and
$\nabla_T$ is the actual temperature gradient in the model. In
practice, the temperature gradient in a WD interior is negligible due
to efficient electron conduction, so this criterion for instability
often reduces to $|B| \gtrsim \nabla_{\rm ad}$ (note that $B<0$ in the
context of phase separation, where the inverted molecular weight
gradient from C enrichment in the mantle acts to excite dynamical
instability). After advancing the crystallization boundary through
zone $k$, we iterate outward starting in zone $k-1$, fully mixing the
contents of $N$ zones $k-1,...,k-N$ until the resulting composition
gradient becomes shallow enough that Eqn~\eqref{eq:Ledoux} is
no longer satisfied in zone $k-N$ and stability is achieved.

The composition adjustments in the preceding steps are made at
constant $P$ and $T$ to maintain hydrostatic equilibrium,
and we use Skye to update the
EOS quantities for affected zones after both phase separation and
mantle mixing composition changes. After these EOS updates, the
crystallized side of the boundary generally moves further into the
solid phase $\phi_k > \phi_{\rm sep}$ due to O enrichment increasing
$\Gamma$ at fixed $T$. We then check whether $\phi_{k-1} >
\phi_{\rm sep}$ so that crystallization must continue advancing through zone
$k-1$ as well. If so, we repeat the preceding steps, separating so
that zone $k-1$ is enriched in O, updating the core boundary
coordinate to encompass $m_{k-1}$, and mixing the C extruded into zone
$k-2$ outward until stability is re-established. We continue iterating
outward and advancing the core boundary through $N_{\rm cr}$ zones until
reaching a point where $\phi_{k-N_{\rm cr}} < \phi_{\rm sep}$ so that zone
$k-N_{\rm cr}$ is outside the crystallization boundary at the end of the
time step. After cooling for another time step, the process of phase
separation will then start again from the location of the saved value of
$m_{\rm cr}$. This ensures that no material that is already in the
solid phase undergoes further separation into an even more O-enriched
composition as long as cooling is monotonic.

\begin{figure}
  \centering
  \includegraphics[width=\apjcolwidth]{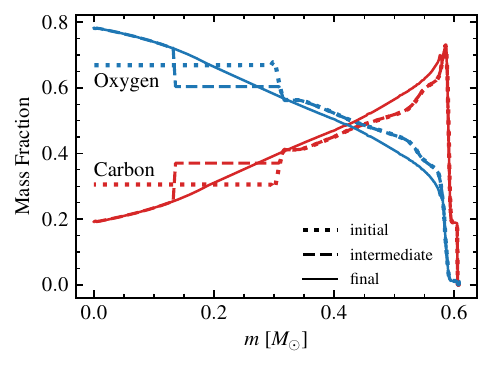}
  \caption{Carbon and oxygen profiles before, during, and after phase
    separation in a 0.6~$M_\odot$ WD model.}
  \label{fig:comp}
\end{figure}

Figure~\ref{fig:comp} shows how the C/O profiles change over the
course of crystallization and phase separation in a MESA model for a
0.6~$M_\odot$ WD using the algorithm described in this section.
As the WD evolves, the outer boundary of the crystallized core
advances outward from the center toward the surface
until eventually encountering the He envelope. We allow phase
separation to proceed until reaching a location where
$X_{\rm C/O} < 0.9$. Once C and O are no longer the dominant elements, we
do not expect pure C/O phase diagrams to be valid, so we do not allow C/O
phase separation to proceed beyond that point. The overall change to
the composition profile seen in Figure~\ref{fig:comp} is similar to
what is seen in other WD codes that implement similar phase diagrams
for C/O crystallization, e.g.\ the LPCODE models of \cite{Althaus2012}
that use the phase diagram of \cite{Horowitz2010}.

\subsection{Phase Separation Heating}

As seen in Figure~\ref{fig:comp}, the net result of phase separation
is to transport some O toward the center of the model while pushing
some C out toward the surface. This mixing changes the
structure and binding energy of the WD interior, resulting in a
heating term that introduces a cooling delay in addition to the latent
heat released by crystallization. We account for this heating term in
the MESA model by making EOS calls that return the internal energy of
each zone before and after the composition changes due to phase
separation during a timestep.
That is, for internal energy $e(\rho,T,\{X_i\})$ from the
Skye EOS, the heating term from phase separation in zone $k$ is calculated over a
timestep $\delta t$ as
\begin{equation}
\epsilon_k = \frac{e(\rho_{k,\rm start},T_k,\{X_i\}_{k,\rm start}) -
  e(\rho_{k,\rm end},T_k,\{X_i\}_{k,\rm end})}{\delta t}~,
\end{equation}
where $\{X_i\}_{k,\rm start}$ is the set of mass fractions
representing the composition of zone $k$ at the
beginning of the step, and $\{X_i\}_{k,\rm end}$ is the composition
after phase separation has modified the composition of zone $k$.
The temperature $T_k$ is held fixed at the start-of-step value during
the phase separation procedure, while $\rho_k$ experiences small
changes due to the composition adjustments at constant pressure $P_k$.
The heating term $\epsilon_k$ from phase separation is then included as a
source term in the energy equation as part of the subsequent stellar
structure solution for the MESA timestep.

\section{White Dwarf Cooling}
\label{s.cooling}

We now demonstrate the effects of our implementation of C/O phase
separation on MESA WD cooling models. In order to make these WD
cooling models and comparisons, we construct two sets of C/O WD starting
models over the mass range $0.5 - 1.0 \, M_\odot$. The first set of
models descends from prior MESA stellar evolution calculations, while
the second set adopts a simple parameterized composition with 50\% C
and 50\% O by mass throughout the WD core for comparison to a common
set of assumptions in the historical literature.
These two sets of models are motivated by the fact that
the best constraints for C/O ratios in WD
interiors remain both theoretically and observationally uncertain.
Historically, the simplest assumption in the face of this uncertainty
has been to adopt a 50/50 C/O ratio (e.g., \citealt{Fontaine2001}),
and many WD phase separation calculations have adopted this
assumption. Stellar evolution models generally produce interior
compositions that are somewhat more rich in O than C, but the exact
ratio is sensitive to reaction rates as well as the details of core
convection during He burning (e.g., \citealt{Straniero2003}). Our
MESA WD models descended from prior stellar evolution calculations
in this work have central O mass fractions of roughly
$0.6-0.7$, which is generally representative of the range of C/O
ratios produced by stellar
evolution codes for standard $^{12}$C$(\alpha,\gamma)^{16}$O
rates. For more detail including composition profiles, see
Appendix~\ref{app:WDmodels}.

We construct our set of WD models descended from full 1D stellar
evolution calculations using the MESA test suite case
{\tt make\_co\_wd}. These models start from the pre-main sequence and evolve
through interior hydrogen and helium burning up to the first thermal
pulse, after which the envelope is stripped to leave a hydrogen-rich
shell of $10^{-3}\, M_\odot$. Residual hydrogen burning and element
diffusion then allow the models to settle onto the WD cooling sequence
as DA WDs with pure hydrogen envelopes of mass
$\approx 10^{-5} - 10^{-4} \, M_\odot$ depending on the underlying WD
mass. We achieve final WD masses in the range $0.5 - 1.0\, M_\odot$
from initial MS masses of $\approx 2.5 - 6.5\, M_\odot$. We save these
starting WD models when they have cooled to a luminosity of
$1\,L_\odot$, near the beginning of the WD cooling sequence.
More information about these WD models can be found in Appendix~\ref{app:WDmodels}.

For comparisons to other codes and phase separation calculations, our
second set of models adopts a simplified interior C/O composition
profile that is precisely 50\% carbon and 50\% oxygen by mass throughout the
interior, with a $10^{-2}\, M_\odot$ He envelope and a $10^{-4}\, M_\odot$
H envelope surrounding the C/O core. We construct these models using
the MESA tool {\tt wd\_builder}%
\footnote{This tool was developed by Josiah Schwab and is now part of
  the MESA-contrib repository (\url{https://github.com/MESAHub/mesa-contrib}).
  It allows constructing MESA WD initial models that specify
  an arbitrary composition and high entropy, rather than constructing
  them through prior stellar evolution.}
to build a $0.6\, M_\odot$ model, and then we rescale to other masses
using the MESA {\tt relax\_mass\_scale} procedure, which scales a model
to a different total mass while keeping the composition profile as a
function of fractional mass coordinate fixed. For simplicity, we set
the parameter {\tt eps\_nuc\_factor} to 0 in these models so that
energy generated by nuclear burning in the hydrogen envelope does not
lead to instability when scaling to higher WD mass, since an H
envelope mass of $10^{-4}\,M_\odot$ is unstable on a WD more massive
than $0.6\,M_\odot$ (e.g., \citealt{Romero2019}).
Without energy generated by nuclear burning, however,
the hydrogen envelope in more massive WDs simply burns down to its
most massive stable configuration at the very beginning of the cooling
sequence, so these WD models have what is often referred to as
``thick'' hydrogen envelopes while avoiding unphysical amounts of
hydrogen for massive WDs.
We refer to these models throughout this paper as ``50/50
C/O'' models.

\subsection{Phase Separation Cooling Delay}

\begin{figure}
  \begin{center}
    \includegraphics{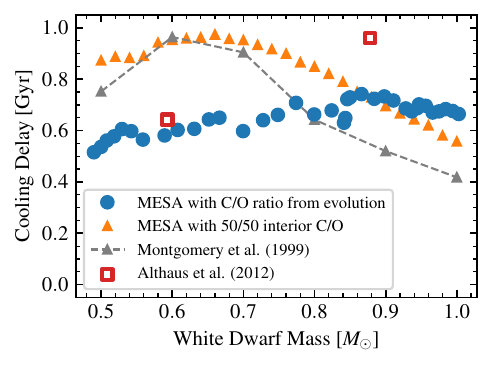}
  \end{center}
  \caption{Cooling delays as a function of WD mass. The gray points
    and curve are from figure 8 of \cite{Montgomery1999} (lower panel,
    curve a), which uses the azeotropic phase diagram of
    \cite{Ichimaru1988}. The red squares are from the models of
    \cite{Althaus2012}, which use the phase diagram of \cite{Horowitz2010}}.
  \label{fig:Delays}
\end{figure}

\begin{figure*}
  \begin{center}
    \includegraphics{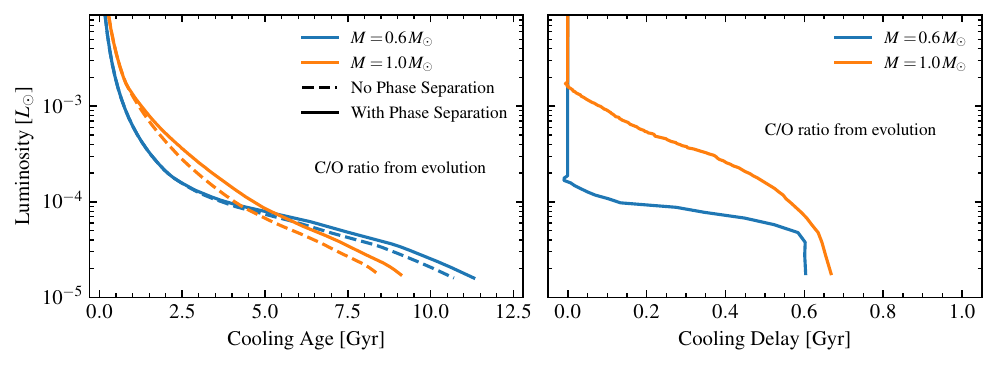}
    \includegraphics{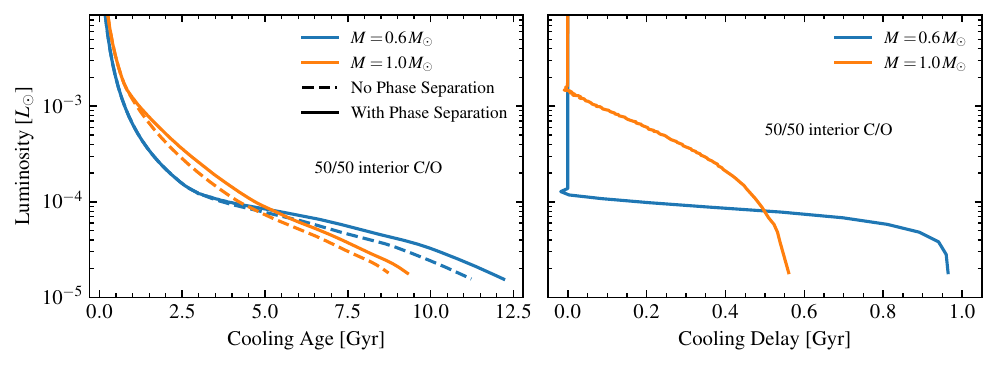}
  \end{center}
  \caption{White dwarf cooling tracks and associated phase separation
    cooling delays for a variety of MESA WD cooling models. The upper
    panels show MESA models with internal composition
    profiles produced by detailed prior evolutionary calculations,
    while the lower panels show MESA models that adopt a simplified
    internal C/O composition that is exactly 50\% C and 50\% O by
    mass.}
  \label{fig:LumTracks}
\end{figure*}

Our WD cooling calculations use these sets of WD starting models,
evolving from an initial luminosity of $1\, L_\odot$ and cooling until
they reach a luminosity of $2 \times 10^{-5}\, L_\odot$. Interior
crystallization and phase separation takes place completely
between these two luminosities for all C/O WD masses considered here,
and so the total difference in cooling age to reach a luminosity of $2
\times 10^{-5} \, L_\odot$ between models with and without phase
separation quantifies the total cooling delay produced by C/O phase
separation.

All our WD cooling models use the atmosphere boundary
conditions of \cite{Rohrmann2012} appropriate for DA WDs, and the
interior thermodynamics are governed by the Skye EOS
\citep{Jermyn2021,Jermyn2023}. This EOS includes the latent heat of
crystallization calculated self-consistently from the local
thermodynamics as described in the previous sections, and this term is
included for all models both with and without phase separation.%
\footnote{Most WD evolutionary codes implement an approximate latent
  heating term of $0.77k_{\rm B}T/\langle A \rangle m_{\rm p}$ based
  on the averaged calculations of \cite{Salaris2000}.
  \cite{Baiko2023} has recently shown that
  the magnitude of the latent heat in C/O mixtures can vary by up to
  $\approx\pm30$\% from the \cite{Salaris2000} value depending on
  C/O ratio. Our models using Skye can dynamically capture this
  variation as the C/O ratio evolves in models experiencing phase
  separation.}
For simplicity in the models in this section, we do not include heavy
element (e.g., $^{22}$Ne) sedimentation, which can introduce an
independent cooling delay of comparable magnitude to that from phase separation
(\citealt{Bildsten2001,Garcia-Berro2008}, see Section~\ref{s.ne22}).
For conductive opacities, all MESA models shown in this section
apply the \cite{Blouin2020apj} correction factors to the
\cite{Cassisi2007} opacities. Appendix~\ref{app:conductive} discusses
some uncertainties in conductive opacities and shows their impact on
the overall cooling delays associated with phase separation in our MESA models.
We run cooling calculations twice for each WD starting model, with and
without phase separation, and the difference in cooling age to reach
the final luminosity therefore represents the total cooling delay produced by
phase separation, independent from sedimentation or latent heat.

Figure~\ref{fig:Delays} shows the total net cooling delay from C/O
phase separation for both of our grids of WD models. For comparison,
the figure also includes the cooling delays calculated by
\cite{Montgomery1999} for WD models with 50/50 C/O interiors when
using the azeotropic phase diagram of \cite{Ichimaru1988}, which is
the most qualitatively similar to our adopted C/O phase diagram
\citep{Blouin2020,Blouin2021} out of the diagrams explored by
\cite{Montgomery1999}. Our set of 50/50 C/O MESA models shows
qualitative agreement with these results of \cite{Montgomery1999} that
use similar composition input and assumptions.
Our models that descend from full evolutionary calculations
show a different trend for cooling delay as a function of
mass, and notably show more variation for small mass changes due to
being descended from different progenitors rather than having smoothly
rescaled overall mass with fixed composition profiles.
Figure~\ref{fig:Delays} also shows delays from
the LPCODE models of \cite{Althaus2012}, which have interior C/O profiles
descended from full evolutionary calculations similar to our MESA
models. These LPCODE models employ the phase diagram of
\cite{Horowitz2010}, which is quite similar to the higher-resolution
phase diagram of \cite{Blouin2020}. Our MESA models therefore yield
very similar overall cooling delays to those seen in the LPCODE, and
the differences seen in Figure~\ref{fig:Delays} are almost entirely
due to our choice of conductive opacities (see
Appendix~\ref{app:conductive} for more detail and MESA models that use
the same conductive opacities as the LPCODE models).

For a typical $\approx 0.6\,M_\odot$ WD, the phase separation cooling delay is
substantially shorter for a model with a C/O composition produced by
prior stellar evolution calculations than for a 50/50 C/O model.
In this case, the details of the phase diagram happen to
maximize the impact of phase separation on cooling timescales at a C/O
ratio around 50/50, while more O-rich C/O ratios result in a somewhat
more modest amount of total separation and cooling delay. So even
though 50/50 C/O would appear to be a simple and agnostic choice of
C/O ratio, it may tend to exaggerate the overall impact of phase
separation on cooling timescales. For more massive C/O WDs, the delays
become more similar due to an overall decreasing trend of the delays
with mass for the 50/50 C/O models.

Figure~\ref{fig:LumTracks} provides more detailed evolutionary tracks
demonstrating the accumulation of the phase separation cooling delay
for $0.6\,M_\odot$ and $1.0\,M_\odot$ models. The upper panels show
delays for models with C/O compositions from evolutionary
calculations, while the lower panels show models with 50/50 C/O
interiors. The $1.0\,M_\odot$ models appear to continue accumulating a
small amount of cooling delay even at very low luminosities, but this
delay is not due to ongoing phase separation. It is rather due to
phase separation having adjusted the interior composition profile
enough that fully crystallized WD models simply have slightly
different thermodynamics and cooling rates even after phase separation
has completed.

\subsection{Comparisons to Other Codes}

\begin{figure}
  \begin{center}
    \includegraphics{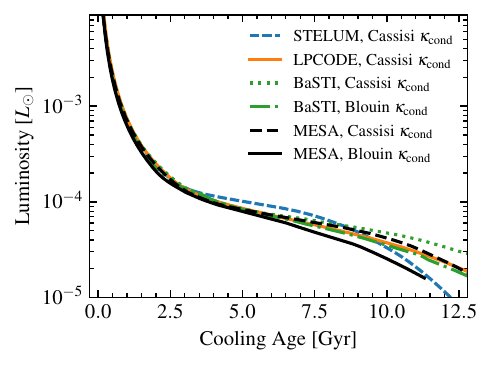}
  \end{center}
  \caption{Comparison of cooling tracks for $0.6\,M_\odot$ DA WD
    models from MESA and three other codes. The STELUM track is from
    \cite{Bedard2020}, the LPCODE track is from \cite{Althaus2015},
    and the BaSTI tracks are from \cite{Salaris2022}.}
  \label{fig:Comparisons}
\end{figure}

As a benchmark to compare our MESA WD cooling implementation to other
commonly used WD cooling tracks, we plot MESA cooling tracks for
$0.6\, M_\odot$ DA WD models in Figure~\ref{fig:Comparisons}. To
facilitate comparisons with other codes that implement alternative
electron conductive opacities, we include both a track that uses only
the \cite{Cassisi2007} conductive opacities as well as a track that
includes the \cite{Blouin2020apj} corrections. We show comparisons to
cooling tracks from the Montreal STELUM ``thick envelope'' model grid
\citep{Bedard2020}, the LPCODE WD models for low-metallicity
progenitors \citep{Althaus2015}, and the BaSTI WD model grid
\citep{Salaris2022}. The STELUM and LPCODE models both use the
uncorrected \cite{Cassisi2007} conductive opacities only.
The BaSTI grid of models provides tracks for the uncorrected
\cite{Cassisi2007} conductive opacities as well tracks for models that
include the \cite{Blouin2020apj} corrections, so we include one of each
for the BaSTI models. In all cases, we have selected the option for
WDs descended from low-metallicity progenitors so that $^{22}$Ne
sedimentation has no impact on the cooling timescales for any of the
models shown in this section ($^{22}$Ne in WD interiors descends from
the primordial CNO abundances of a WD progenitor star).

The STELUM models \citep{Bedard2020,Bedard2022} are very widely used as a
reference for WD cooling timescales, but the widely used model tracks
that are published online%
\footnote{\url{https://www.astro.umontreal.ca/~bergeron/CoolingModels/}}
make a number of simplifying
assumptions that are not ideal for accurate WD cooling at late times.
Unlike the other models in this section, the STELUM models use gray
Eddington atmosphere boundary conditions rather than tabulated outer
boundaries that account for non-gray radiative transfer such as
those of \cite{Rohrmann2012}. These models also assume flat 50/50
interior C/O ratio by mass, and while STELUM does include
options for C/O phase separation \citep{Bedard2022}, the public tracks
based on \cite{Bedard2020} do not include any C/O phase
separation upon crystallization. Therefore, the STELUM cooling track
in Figure~\ref{fig:Comparisons} does not agree well with MESA or the
other WD codes at low luminosities and temperatures. MESA models can closely
recover the STELUM cooling track by selecting options to emulate all of the
above simplifying assumptions.

The LPCODE model \citep{Althaus2015}%
\footnote{This paper includes models from several different
  progenitors with low metallicity, and we select the track from the 
  $Z=3 \times 10^{-5}$ model for the comparison here.
  The other models in \cite{Althaus2015} have a
  significant extra cooling delay at earlier times due to residual
  hydrogen burning, but that effect is negligible in the $Z=3 \times
  10^{-5}$ model.}
in Figure~\ref{fig:Comparisons} includes very similar
input physics assumptions to the MESA model: they both use the same
tabulated DA atmosphere boundary conditions \citep{Rohrmann2012} and
very similar interior C/O composition profiles produced by
prior stellar evolution models. The LPCODE model
employs the phase diagram of \cite{Horowitz2010} for crystallization
and phase separation, which  is very similar to the
\cite{Blouin2021} phase diagram employed by MESA \citep{Blouin2020}.
Reassuringly, the MESA model that makes the same
assumption about conductive opacities matches the LPCODE cooling track
very closely.

\begin{figure*}
  \begin{center}
    \includegraphics{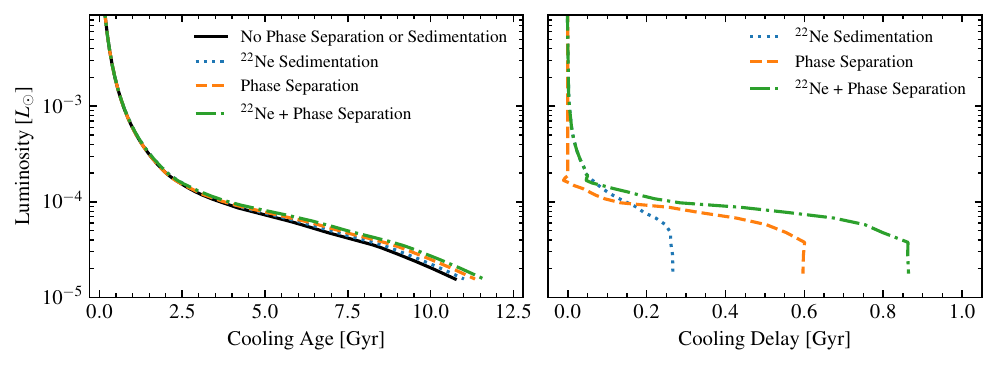}
  \end{center}
  \caption{WD cooling delays associated with both phase separation and
  $^{22}$Ne sedimentation for a $0.6\,M_\odot$ DA WD model.}
  \label{fig:Ne22Tracks}
\end{figure*}

The BaSTI models \citep{Salaris2022} make very similar assumptions to
the models of both MESA and LPCODE in terms of atmosphere boundary
conditions, interior composition profiles, phase diagram, and phase
separation. However, they exhibit much slower cooling after
crystallization, and we speculate that this is due to differences in the
thermodynamics of the solid phase reported by their choice of EOS.

\section{Relation to Neon Sedimentation}
\label{s.ne22}

In degenerate WD interiors, hydrostatic equilibrium establishes an
electric field that approximately balances gravity
for the dominant ion species C and O \citep{Bildsten2001,Chang2010}.
The net force experienced by ion species $i$ is
$F_i = - A_i m_{\rm p} g + Z_i eE$, with $eE \approx 2 m_{\rm p} g$,
which approximately cancels for the dominant background, while ions
with extra neutrons ($A_i > 2Z_i$) such as $^{22}$Ne and $^{23}$Na
experience a net downward force that can lead to sedimentation
\citep{Bauer2020}.
$^{22}$Ne in particular is present in WD interiors at a mass fraction
that reflects the initial metallicity $Z$ of their progenitor stars,
as CNO burning results in most of the core metallicity becoming
$^{14}$N by the end of the main sequence, which then burns to
$^{22}$Ne during the He burning that forms the C/O WD core.

Heavy-element sedimentation causes heating that can slow WD
cooling \citep{Isern1991,Bildsten2001,Deloye2002}, with the magnitude of this
effect scaling up with WD progenitor metallicity, and this has been
implemented and studied in several WD evolution codes
\citep{Garcia-Berro2008,Paxton2018,Bauer2020,Salaris2022}.
This cooling delay operates in addition to any delay from
crystallization and phase separation, and in particular the confluence
of these two delays has been invoked to explain the slow cooling of
the WD population in the metal-rich ($Z\approx 0.04$) open cluster NGC~6791
\citep{Garcia-Berro2010,Althaus2010}. In this section we consider MESA
models in which both phase separation and heavy-element sedimentation
processes operate, and compare to the delays seen in other WD codes.

Our models in this section all descend from $Z=0.02$ progenitors, so
that the interior mass fraction of $^{22}$Ne is roughly 0.02
throughout the C/O core. We include element diffusion in the liquid
portion of the WD core following the treatment described in
\cite{Paxton2018,Bauer2020,Jermyn2023}, with a smooth turnoff of
diffusion in the crystallized solid regions according to the smoothed
phase parameter $\phi$ introduced in previous sections.
The diffusion coefficients in the strongly-coupled plasma of the WD
core are those of \cite{Caplan2022}, which closely match the
coefficients of \cite{Hughto2010} that are often used for $^{22}$Ne
sedimentation in WDs.

Figure~\ref{fig:Ne22Tracks} shows cooling tracks and associated
cooling delays for models with and without both $^{22}$Ne
sedimentation heating and C/O phase separation. Our baseline model
includes element diffusion but not phase separation, and omits the
heating term associated with heavy-element sedimentation. Our second
model turns on this heating term, which is responsible for the modest
$\approx 0.3\,\rm Gyr$ delay seen in the dotted blue curve. Our third
model turns this sedimentation heating back off while turning on phase
separation, resulting in a somewhat more substantial delay of $\approx
0.6\,\rm Gyr$ seen in the dashed orange curve. Our final model turns
on both of these effects together to yield a total delay of
$\approx 0.9\,\rm Gyr$. This verifies that our implementations of both
these effects in MESA are compatible with each other and lead to the
expected total delay from both independent processes.

With our implementation of sedimentation including state-of-the-art
diffusion coefficients \citep{Caplan2022}, our WD cooling delays from
$^{22}$Ne sedimentation are significantly shorter than those seen in
LPCODE models of C/O WDs at the same metallicity \citep{Camisassa2016},
regardless of whether C/O phase separation is considered.
This is despite having implementations of other WD cooling physics
including crystallization and phase separation that very closely match
other aspects of the LPCODE models (see Figure~\ref{fig:Comparisons}).
LPCODE models of C/O WDs at $Z=0.02$ exhibit $^{22}$Ne cooling delays
greater than $1\,\rm Gyr$ according to \cite{Camisassa2016}. In
contrast, our MESA models show a delay of only $0.3\,\rm Gyr$, and
recent BaSTI models that include $^{22}$Ne sedimentation also show a
relatively modest delay of $\approx 0.5\,\rm Gyr$ for $0.6\,M_\odot$
WDs descended from $Z=0.02$ progenitors \citep{Salaris2022}. The
original implementation of $^{22}$Ne sedimentation in LPCODE from
\cite{Garcia-Berro2008} also showed much more modest cooling delays
when employing their default $^{22}$Ne diffusion coefficient based on
\cite{Deloye2002}. While more recent MD simulations have yielded
better constraints on the diffusion coefficient
\citep{Hughto2010,Caplan2022}, the overall change relative to the
coefficient employed by \cite{Garcia-Berro2008} is modest and should
only result in slightly slower sedimentation, so it is not clear why more
recent LPCODE models show much larger $^{22}$Ne sedimentation delays
\citep{Althaus2010,Camisassa2016,Camisassa2021}.

We therefore argue that the $^{22}$Ne cooling delays in LPCODE models
may be significantly overestimated, and this could have important implications
for some inferences based on the cooling timescales of these
models. LPCODE models have been used to resolve a discrepancy
in the WD cooling ages for the metal-rich open cluster NGC~6791, which
was originally thought to have a main sequence turnoff age more than
$2\,\rm Gyr$ older than the age inferred from comparing the WD luminosity
function to WD cooling models \citep{Bedin2008}.
\cite{Garcia-Berro2010} used LPCODE models to argue that the inclusion
of both C/O phase separation and $^{22}$Ne sedimentation could provide a
long enough cooling delay to resolve this discrepancy. However, the
faster cooling timescales and more modest $^{22}$Ne delays seen in our
MESA models would reintroduce tension between the age inferred from
the WD luminosity function and the ages inferred from the other
stellar populations in NGC~6791.

\cite{Camisassa2021} have also
recently used LPCODE models for WDs descended from very metal-rich
progenitors ($Z=0.06$) to argue that $^{22}$Ne sedimentation could
explain the anomalous observed kinematics for some $Q$-branch WDs,
interpreted as evidence of an $\approx 8\,\rm Gyr$ cooling delay
for some massive WDs \citep{Cheng2019}. However, since MESA models
show less than half the cooling delay compared to LPCODE models for a
given amount of $^{22}$Ne sedimentation, an explanation invoking
standard sedimentation alone would require an implausibly high
$^{22}$Ne abundance corresponding to $Z \gtrsim 0.15$ if our MESA
models are accurate.

While heavy-element sedimentation alone is likely insufficient to
explain the cooling delays in either NGC~6791 or the $Q$-branch WDs,
the recently proposed process of $^{22}$Ne ``distillation''
\citep{Blouin2021apjl} is a promising avenue for explaining these
delays by enhancing transport of $^{22}$Ne toward the center through
dynamical mixing. This process is somewhat analogous to C/O phase
separation in that crystallization of a multi-component plasma induces
some amount of instability and mixing, but the physics is more
complicated due to requiring a three-component phase diagram rather
than the simpler two-component diagram employed for C/O phase
separation in this work. Still, it appears that distillation shows the
most promise for explaining cooling delays in WDs descended from
metal-rich ($Z \gtrsim 0.02$) populations, and may in fact be required
by observations such as those of NGC~6791 and the $Q$-branch. Our
present work with MESA lays the groundwork for future investigation of
distillation based on three-component phase diagrams
\citep{Blouin2021apjl,Caplan2023} in full WD evolutionary models.

\section{Conclusions}

We have provided a new implementation of C/O phase separation during
WD crystallization as part of the publicly available stellar evolution
software MESA. Our implementation relies on the phase diagram of
\cite{Blouin2021}, and produces a net transport of O toward the center
and C toward the surface. This rearrangement of the interior chemical
profile during WD crystallization provides an energy source that can
delay WD cooling. In our detailed MESA models, the net cooling delay
is typically on the order of $0.5\, \rm Gyr$. This is somewhat smaller
than often quoted delays of $\approx 1\,\rm Gyr$, which can arise from
older phase diagrams that led to more dramatic separation in typical
WD models, or from models that assume core C and O mass fractions of
exactly 0.5, where the phase diagram happens to maximize the impact of
separation. Our WD models descended from MESA models of prior stellar
evolution have central O mass fractions of roughly $0.6 - 0.7$ depending on mass.

Our models that include C/O phase separation
show good agreement with other WD models that implement
comparable physics, such as the LPCODE models of \cite{Althaus2012,Althaus2015}.
However, there is tension in the overall WD cooling timescales between
MESA and LPCODE WD cooling models that implement comparable physics
with both C/O phase separation and heavy-element sedimentation for WDs
descended from metal-rich ($Z\gtrsim 0.02$) progenitor populations.
If the cooling timescales of our MESA models are correct, this may
reintroduce discrepancies between WDs associated with observed
metal-rich populations (e.g., NGC~6791 and the $Q$-branch) and
state-of-the-art WD cooling models that include the most up-to-date
input physics. This would then require the implementation of newly
proposed pieces of input physics to resolve these discrepancies, such
as distillation in three-component mixtures
\citep{Blouin2021apjl}. Our progress with MESA in this work lays the
groundwork for future investigations to explore these new physical
processes with detailed WD evolution models.

\vspace{1em}
{\it Acknowledgments:}
  EB gratefully acknowledges helpful conversations with Simon Blouin,
  Mar\'ia Camisassa, Charlie Conroy, Bart Dunlap, Mike Montgomery, and Alejandra
  Romero during the development of this work. EB is also grateful to
  the anonymous referee for helpful suggestions improving both the
  text and figures for the key results of this work.
  EB also thanks the MESA developers team for building and maintaining
  the open software upon which this work rests, especially Adam Jermyn
  and Josiah Schwab for developing the particular capabilities
  leveraged by this work.
  This work benefited from discussions at the KITP program
  {\it White Dwarfs as Probes of the Evolution of Planets, Stars, the
    Milky Way and the Expanding Universe} in the Fall of 2022,
  and was supported in part by the National Science Foundation under
  Grant No.\ NSF PHY-1748958.

\software{%
  \texttt{MESA} \citep{Paxton2011, Paxton2013, Paxton2015, Paxton2018,
    Paxton2019, Jermyn2023},
  \texttt{Skye} \citep{Jermyn2021},
  \texttt{matplotlib} \citep{hunter_2007_aa},
  \texttt{NumPy} \citep{der_walt_2011_aa}, and
  \texttt{Python} from \href{https://www.python.org}{python.org}.
}

\clearpage
\appendix

\section{White Dwarf Models}
\label{app:WDmodels}

\begin{figure*}
  \begin{center}
    \includegraphics{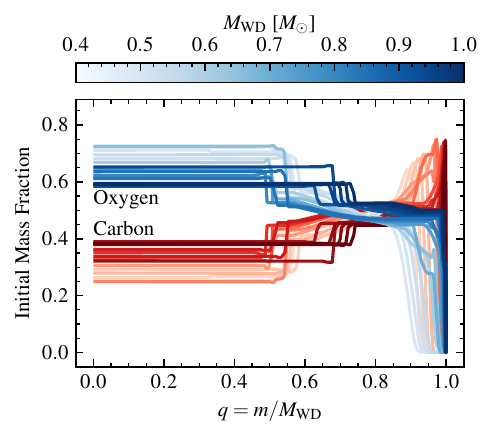}
    \includegraphics{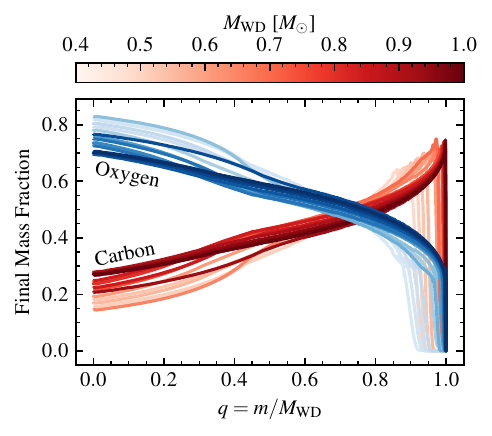}
  \end{center}
  \caption{Carbon and oxygen composition profiles versus fractional
    mass coordinate
    $q=m/M_{\rm WD}$ for our MESA WD models with masses in the range
    $0.5 - 1.0 \, M_\odot$, before (left) and after (right) C/O phase
    separation. Darker line colors indicate more massive WD models.
    For the right panel, ``after'' phase separation is defined
    as having reached a point where the C/O-dominated layers have
    completely crystallized, so that no further mixing due to phase
    separation will occur.
  }
  \label{fig:compositions}
\end{figure*}

Figure~\ref{fig:compositions} shows the C/O composition profiles from
a representative subset of our WD models descended from full prior stellar
evolution calculations. We use MESA to evolve our models from ZAMS
through H and He burning up to the point of the first thermal pulse on
the AGB. Once the first thermal pulse occurs, we artificially strip
the outer envelope and turn on diffusion to allow the models to settle
onto the WD cooling sequence with pure H atmospheres as DA WDs. Since
this step of the process circumvents realistic AGB mass loss, it does
not produce an accurate initial-to-final mass relation (IFMR), but it
does produce a representative range of C/O profiles in the interior
core region where burning has ceased, so this is sufficient for our
study in this work. The starting interior C/O ratio produced by
evolution prior to the WD stage is sensitive to the
$^{12}$C$(\alpha,\gamma)^{16}$O reaction rate
\citep{Fields2016,deBoer2017,Chidester2022}. We adopt the \cite{Kunz2002}
rate for this reaction for our models in this work, which is
quite similar to the \cite{deBoer2017} rate for the temperature range
relevant for He core and shell burning to produce WD interior C/O
profiles. Our MESA WD models show core O mass fractions in the range
$\approx 0.6-0.75$, with the lower-mass models typically on the more
O-rich end in their interiors. The low-mass models also tend to have
very thick He envelopes of up to about $0.05\,M_\odot$, but the He
envelope masses are thinner than $10^{-2}\,M_\odot$ for typical WD
masses around $0.6\,M_\odot$, and as thin as $10^{-3}\,M_\odot$ for
the most massive WDs around $1.0\,M_\odot$.

While the interior C/O composition profiles of our MESA models are
generally representative of the current state of the art for 1D
stellar evolution models, we feel it is also worth noting that
significant uncertainty remains as to the precise interior composition and
especially the ratio of C to O in WD cores. Helium burning generically
coincides with $\alpha$ captures onto $^{12}$C to produce a mixture of
C and O, but uncertainty in the $^{12}$C$(\alpha,\gamma)^{16}$O
reaction rate makes it difficult to predict the precise ratios from 1D
stellar evolution calculations
\citep{deBoer2017,DeGeronimo2017,DeGeronimo2019,Chidester2022}.
It has been suggested that WD asteroseismology could be used to
constrain interior structure and composition of WD models, and
ultimately may even yield constraints on the
$^{12}$C$(\alpha,\gamma)^{16}$O reaction rate as a result
\citep{Metcalfe2001,Metcalfe2002,Fontaine2002,Metcalfe2003}.
Indeed, some recent studies have attempted to use seismology to constrain WD
interior compositions \citep{Giammichele2018,Charpinet2019,Giammichele2022},
yielding central O mass fractions as high as 0.86, which would be difficult
to produce with reasonable assumptions in 1D stellar evolution models
\citep{DeGeronimo2019}. However,
\cite{Timmes2018} and \cite{Chidester2021} have pointed out some
additional pieces of physics that are not accounted for in the
parameterized models used for the \cite{Giammichele2018} fitting
procedure, and including these extra effects on the structure can
shift the mode frequencies beyond the quoted uncertainties of the models.
\cite{Bell2022} also recently showed that the
\cite{Giammichele2018,Charpinet2019} fits for the WD KIC~08626021 are
inconsistent with the astrometry from {\it Gaia}, suggesting that the
seismological fitting procedure may be overly flexible and yield
much larger uncertainties than currently appreciated. Some other
recent WD seismology studies have pointed out that pulsation spectra
are far more sensitive to other aspects of WD structure such as
composition transition locations and shapes, making it difficult to
confidently isolate a pulsation signal firmly constraining the central
C/O mass fractions (e.g., \citealt{Bischoff-Kim2019,Bell2019}).

\section{Conductive Opacities}
\label{app:conductive}

\begin{figure}
  \begin{center}
    \includegraphics{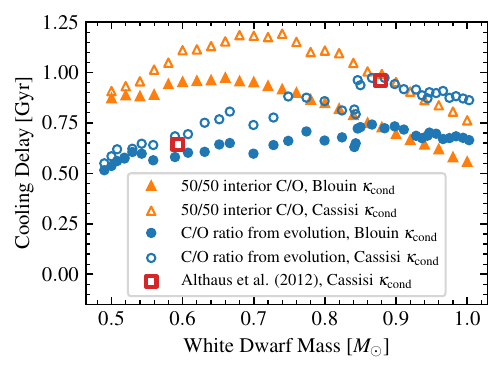}
  \end{center}
  \caption{Cooling delays as a function of WD mass, similar to
    Figure~\ref{fig:Delays}, but including models with different
    conductive opacities. Open symbols represent models using only the
    conductive opacities of \cite{Cassisi2007}, while filled symbols
    represent models that include the corrections of
    \cite{Blouin2020apj}.}
  \label{fig:Delays_appendix}
\end{figure}

The current default treatment of conductive opacities in MESA applies
correction factors to the classic values of \cite{Cassisi2007} based
on the recent calculations of \cite{Blouin2020apj}. However,
\cite{Blouin2020apj}, \cite{Cassisi2021}, and \cite{Salaris2022} have
all pointed out that there is still substantial uncertainty in how
best to bridge between the regimes of moderate and strong electron
degeneracy for conductive opacities. 
WD models typically cool faster at late times when applying the
\cite{Blouin2020apj} conductive opacity corrections, which can decrease
the effective opacity by a factor of 2-3 in some regimes
\citep{Blouin2020apj, Salaris2022, Jermyn2023}. Figure~\ref{fig:Delays_appendix}
shows phase separation cooling delays for MESA models that use the
\cite{Cassisi2007} opacities without any corrections, alongside the
models from Section~\ref{s.cooling} that include the corrections from
\cite{Blouin2020apj}. While the
qualitative trends are similar in either case, the overall cooling
delays associated with crystallization and phase separation are
somewhat shorter when applying the \cite{Blouin2020apj} opacities
because of the overall faster cooling for these models.
The LPCODE models from \cite{Althaus2012} employ the
\cite{Cassisi2007} conductive opacities as well, and agree very well
with our models that use the same conductive opacities.
As the focus of the current work is simply to describe our implementation of
phase separation and quantify its effects, we withhold judgment on
which set of conductive opacities offers more accurate absolute
WD cooling timescales, and emphasize that MESA makes either set of
opacities easily available to those who wish to explore further.

\bibliography{refs}
\bibliographystyle{yahapj}

\end{document}